\documentclass[a4paper,11pt]{article}
\pdfoutput=1 

\usepackage{jheppub} 

\usepackage{bm,amsmath,amssymb,slashed,graphicx,caption,%
            enumerate,alltt,xspace,multirow,xcolor,mathrsfs}

\makeatletter
\g@addto@macro\bfseries{\boldmath}
\makeatother

\usepackage{listings}
\usepackage{bbold}

\newcommand{\POWHEGBOX}{{\tt POWHEG BOX}}
\newcommand\mathd{\mathrm{d}}

\newcommand{\bz}{b_0}
\newcommand{\nf}{n_{\rm f}}

\newcommand{\as}{\alpha_s}

\newcommand{\MSbar}{\ensuremath{\overline{\text{MS}}}}
\newcommand{\MSB}{\MSbar}
\newcommand{\cF}{C_d}
\newcommand{\muF}{\mu_{\scriptscriptstyle \rm F}}

\newcommand{\pt}{p_{\rm \scriptscriptstyle T}}
\newcommand{\ptz}{p_{\rm \scriptscriptstyle T,Z}}
\newcommand{\yz}{y_{\rm \scriptscriptstyle Z}}
\newcommand{\ptcut}{p_{\rm \scriptscriptstyle T,cut}}
 
\newcommand\kz{k_{\scriptscriptstyle Z}}
\newcommand\kq{k_{q}}
\newcommand\kd{k_{d}}
\newcommand\kbarq{k_{\bar q}}
\newcommand\kg{k_{g}}
\newcommand\kgstar{k_{g^*}}

\newcommand\veckz{\vec{k}_{\scriptscriptstyle Z}}

\newcommand\veckd{\vec{k}_{d}}

\newcommand{\PhiB}{{\bf \Phi}_{\rm\scriptscriptstyle B}}
\newcommand{\phiB}{\Phi_{\rm\scriptscriptstyle B}}
\newcommand{\phiD}{\Phi_{\rm\scriptscriptstyle D}}
\newcommand{\Phig}{{\bf\Phi}_{g}}
\newcommand{\phig}{{\Phi_{g}}}
\newcommand{\Phiqqb}{{\bf\Phi}_{q\bar{q}}}
\newcommand{\phiqqb}{\Phi_{q\bar{q}}}
\newcommand{\Phigstar}{{\bf\Phi}_{g^*}}
\newcommand{\Phiplus}{{\bf\Phi}_{\scriptscriptstyle \oplus}}
\newcommand{\Phiminus}{{\bf\Phi}_{\scriptscriptstyle\ominus}}
\newcommand{\splus}{C_{\scriptscriptstyle\oplus}}
\newcommand{\sminus}{C_{\scriptscriptstyle\ominus}}

\newcommand{\Bqqb}{{\cal D}_{g}}
\newcommand{\Bqqbstar}{{\cal D}_{g^*}}
\newcommand{\Rqqb}{{\cal D}_{q\bar{q}}}

\newcommand{\sB}{B}
\newcommand{\sg}{R_{g}}
\newcommand{\sgstar}{R_{g^*}}
\newcommand{\sV}{V}
\newcommand{\sqqb}{R_{q\bar{q}}}

\newcommand\sss{\mathchoice%
{\displaystyle}%
{\scriptstyle}%
{\scriptscriptstyle}%
{\scriptscriptstyle}%
}

\newcommand\nplus{\oplus}
\newcommand\nminus{\ominus}

\newcommand\sssplus{{\sss \nplus}}
\newcommand\sssminus{{\sss \nminus}}

\newcommand\xplus{x_\sssplus}
\newcommand\xminus{x_\sssminus}
\newcommand\pplus{p_\sssplus}
\newcommand\pminus{p_\sssminus}

\newdimen\hbigcirc
\newdimen\wbigcirc

\newcommand{\muC}{\mu_C}
\newcommand{\TF}{T_{\rm F}}

\newcommand\qi{q}
\newcommand\qbi{{\bar q}}
\newcommand\kqqb{k_{\qi\qbi}}

\title{Infrared Renormalons in Kinematic Distributions for Hadron Collider Processes}
\preprint{
  \begin{flushright}
    OUTP-20-13P, IPPP/20/60
  \end{flushright}
}

\author[a,b]{Silvia Ferrario Ravasio,}
\author[c]{Giovanni Limatola,}
\author[c]{Paolo Nason}

\emailAdd{silvia.ferrarioravasio@physics.ox.ac.uk}
\emailAdd{g.limatola@campus.unimib.it}
\emailAdd{paolo.nason@mib.infn.it}

\affiliation[a]{Institute for Particle Physics Phenomenology, University of Durham,  Durham DH1 3LE, UK}
\affiliation[b]{Rudolf Peierls Centre for Theoretical Physics, University of Oxford, Clarendon Laboratory, Parks Road, Oxford OX1 3PU, UK}
\affiliation[c]{Universit\`a di Milano-Bicocca and INFN, Sezione di
  Milano-Bicocca, Piazza della Scienza 3,20126 Milano, Italy}

\date{Received: date / Accepted: \today}

\abstract{
  Infrared renormalons in Quantum Chromodynamics
  are associated with non-perturbative corrections to
  short distance observables. Linear renormalons, i.e. such
  that the associated non-perturbative corrections scale like one inverse power of
  the hard scale, can affect at a non-negligible level
  even the very high-energy phenomena studied at the Large Hadron Collider.
  Using an Abelian model, we study the presence of linear
  renormalons in the transverse momentum distribution of a neutral
  vector boson $Z$ produced in hadronic collisions. We consider a process
  where the $Z$ transverse momentum is balanced by a sizable recoil against a
  coloured final state particle. One may worry that such a colour configuration,
  not being azimuthally symmetric, could generate unbalanced soft radiation,
  associated in turn with linear infrared renormalons affecting
  the transverse momentum distribution of the vector boson.
  We performed a numerical calculation of the
  renormalon effects for this process in the so-called large $b_0$ limit.
  We found no evidence of linear renormalons in the transverse
  momentum distribution of the $Z$ in the large transverse-momentum region,
  irrespective of rapidity cuts.
}


\begin{document}


\maketitle


\newcommand{\citere}[1]{Ref.\,\cite{#1}}
\newcommand{\citeres}[1]{Refs.\,\cite{#1}}

\section{Introduction}
\label{sec:intro}
Due to the absence of clear new physics signals at the Large Hadron
Collider (LHC), precision calculations for LHC-physics processes have
become increasingly important in recent years, in order to aid in the search for new
physics phenomena that manifest themselves as modest deviations of the production and decay
properties of the Standard Model particles.
Improving the precision of QCD calculations is
particularly demanding, given the size of the strong coupling constant,
and a considerable effort is under way to improve the precision of QCD calculations
of collider processes with the inclusion of two-loop, and even
three-loop corrections.

The transverse momentum spectrum of the $Z$ boson is one of the most
precise observables measured at the LHC.  More specifically, the
normalized distributions are measured with a precision that reaches
the sub-percent level in the low-intermediate values of the transverse
momentum~\cite{Khachatryan:2015oaa,Aad:2015auj,Sirunyan:2019bzr,Aad:2019wmn}.
The uncertainties associated with theoretical calculations are,
however, still at the percent level.  The process of $Z$+jet
production in hadronic collisions has been computed at NNLO
QCD~\cite{Boughezal:2016, Gehrmann:2016, Gehrmann:2018}.  The current
state of the art is given by NNLO+N${}^3$LL~\cite{Bizon:2019zgf}, and
the resummation effects have been proven to have a large effect in the
region where the $Z$ boson transverse momentum is small, which is also
plagued by important non perturbative
effects~\cite{Dokshitzer:1978yd,Parisi:1979se,Collins:1984kg}.
The ATLAS measurement at
13~TeV~\cite{Aad:2019wmn} is in excellent agreement with the
NNLO+N${}^3$LL result for $p_{\rm \scriptscriptstyle T}^{Z}<30$~GeV,
while some tension at the level of few percent is found for larger
values, where the impact of resummation is negligible. This tension
is similar in size to the residual scale uncertainty, and, furthermore,
it is not observed in the 8~TeV data~\cite{Bizon:2018foh}.
These measurements may have important implications for constraining
the strong coupling constant and the PDFs at the LHC
(see e.g. Ref.~\cite{Boughezal:2017nla}).

In this article we investigate the presence of non-perturbative effects
in the $Z$-boson transverse-momentum distribution in the moderately
large transverse momentum region,
where resummation effects should not play an important role.
We consider both the single and double differential cross sections
\begin{equation}
  \frac{\mathd \sigma}{\mathd^2 p_T},\quad \frac{\mathd \sigma}{\mathd^2 p_T\mathd y},
\end{equation}
where $p_T$ and $y$ are the transverse momentum and rapidity of the $Z$ boson,
compute explicitly their cumulants $\sigma(p_T>p_{\rm cut})$,
$\sigma(p_T>p_{\rm cut},0<y<y_{\rm cut})$, and look for linear
renormalons in the result.

If we assume that this distribution is affected by linear non-perturbative corrections,
their natural size would be of order $\Lambda/\pt$,
where $\Lambda$ is a typical hadronic scale and
$\pt$ is the $Z$ transverse momentum, thus leading to values that may easily reach the few percent level,
larger than the present experimental error
and the current theoretical accuracy.
A further concern stems from
the fact that the soft-radiation pattern in the $Z+$jet process is not azimuthally
symmetric, and infrared renormalons are related to soft radiation. If we believe that
we can model renormalon effects as being equivalent to the emission of a soft
particle with transverse momentum of order $\Lambda$, it would be reasonable to
assume that they may affect linearly the transverse momentum of the recoil system,
and thus also the transverse momentum of the $Z$ boson.

There is a well-known relation between non-perturbative corrections
and the factorial growth of the coefficients of the perturbative
expansion in field theories.
An observable $R$, in a generic renormalizable quantum field theory,
can be expressed as a series in the renormalized coupling $\alpha$
\begin{equation}
\label{eqn:series}
R(\alpha)=\sum_n r_n\alpha^n.
\end{equation}
Generally, the coefficients $r_n$ grow factorially for large orders,
and one must worry about the associated ambiguities in the definition of the sum
in~\eqref{eqn:series}. 
There are three known sources of factorial growth: ultraviolet (UV) renormalons,
infrared (IR) renormalons and instantons. In QCD, IR renormalons lead to
a factorial growth
of the form
\begin{equation}
  r_n\propto (2b_0/p)^n n! \label{eq:RenormalonGrowth}
\end{equation}
where $p$ is a positive integer, $b_0$ is defined as usual
\begin{equation}
  b_0=\frac{33-2n_l}{12\pi},
\end{equation}
and $n_l$ is the number of light flavours. Eq.~(\ref{eq:RenormalonGrowth}) implies that
the terms of the perturbative expansion grow by a factor $n (2b_0/p) \as$ as the order increases,
and thus reach a minimum when $n \approx p/(2b_0\as)$.
Using Stirling's formula ($n!\approx n^n\exp(-n)$),
the size of the minimal term is easily estimated to be
\begin{equation}\label{eq:RenormPower}
  \exp\left(-\frac{p}{2b_0\as}\right) \approx  \exp\left(-\frac{p}{2} \log\frac{\mu^2}{\Lambda^2}\right)
  \approx \left(\frac{\Lambda}{\mu}\right)^p,
\end{equation}
where $\mu$ is the renormalization scale.
This is the well known connection between IR renormalons and power corrections. It tells us that
the size of the minimal term of the perturbative expansion plays the role of a power suppressed
ambiguity in the value of the corresponding observables. In our case, as already mentioned earlier,
we worry about the case when $p=1$, that will be referred to in the following as \emph{linear renormalons}.
We also stress that, due to the relatively large value of the QCD
coupling constant, the minimal term in the perturbative expansion may be reached
quite early, and one may worry that, in some cases, already at the N$^3$LO level
linear renormalons may become relevant.

UV renormalons also lead to a factorial growth of the same form as in eq.~(\ref{eq:RenormalonGrowth}), with $p$ assuming negative integer values starting with $-2$.
In this case, formula~(\ref{eq:RenormPower}) still holds if we replace $p$ with $|p|$,
and the minimal term is, at worse, of order $\Lambda^2/\mu^2$.
Furthermore, the perturbative expansion is alternating in sign, and can
in principle be resummed with Borel techniques.
Instantons lead instead to a much stronger power suppression,
and need not be considered here.\footnote{Renormalons have been first discussed in the 1970s
  in references \cite{Gross:1974,Lautrup:1977,tHooft:1977xjm},
  and have received attention from
  a phenomenological perspective since the 1992 in refs.~\cite{Brown:1992,ZAKHAROV1992452,Mueller:1984vh}. 
  A well-known review of this topic is given in~\cite{Beneke:1998ui}.}

General arguments based upon field theory assure us that
linear renormalons cannot be present in certain quantities. This is
the case for observables that admit an Operator Product Expansion (OPE), and
are such that there are no operators of dimension higher
by one power with respect to the leading contribution. In fact, IR
renormalons in hard processes originate from subgraphs with low
momenta, and, if the process in question admits an OPE, it must be
possible to organize the Feynman graphs in terms of expectation values of local
operators~\cite{Parisi:1978az,Mueller:1984vh}.  In case of processes that do not
have a known OPE, we lack a safe guideline for the classification of
non-perturbative corrections, and thus also of renormalon effects. In such cases,
one has to resort to assumptions and approximations in order to
gain some insight into the problem. 

A very valuable method for the study of IR renormalons is the so-called large-$\bz$ approximation (see \cite{Beneke:1998ui} and references therein).
In essence, the method considers
an alternative theory such that the renormalon structure can
be fully computed, i.e. an Abelian gauge theory in the limit of a
large number of fermions $n_f$.  In this theory the leading terms, of
order $(\as \nf)^k$, are fully calculable, and the theory
exhibits IR and UV renormalons. In order to have an estimate of
the renormalon effects in the full original theory, at the end of the calculation one
replaces the $\bz$ in the Abelian theory (that is proportional
to $-\nf$) with the full QCD $\bz$.

We recall that the authors of ref.~\cite{Beneke:1995pq}
showed that, in the large-$\bz$ approximation, the Drell-Yan cross section
does not have linear renormalons. This result was used to argue that,
contrary to previous claims, soft gluon resummation cannot give
solid evidence of the presence of linear renormalons in the total
Drell-Yan cross section. Later on, in ref.~\cite{Dasgupta:1999zm}, it was shown
that under certain assumptions the rapidity distribution of Drell-Yan
pairs is free of linear power corrections. This result can be shown to imply
that in the large-$\bz$ approximation there are no linear renormalons
in the rapidity distribution of Drell-Yan pairs.

\begin{figure}[tb]
  \centering
  \includegraphics[width=0.6\linewidth]{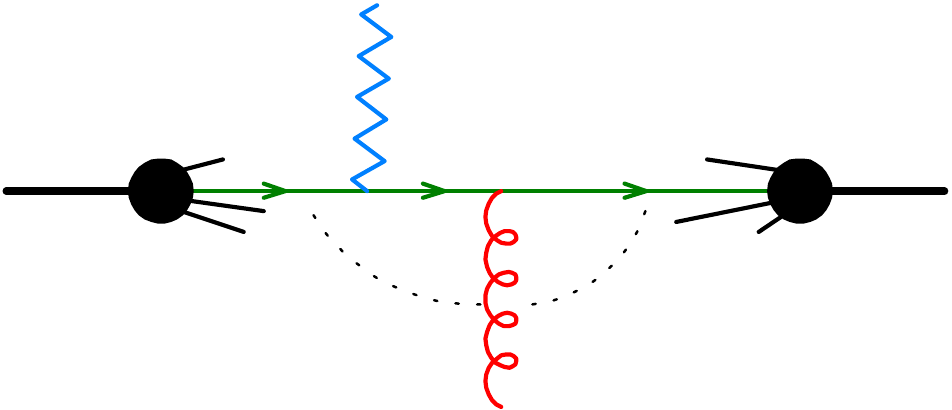}
  \caption{\label{fig:SoftPatternZ} Typical diagram contributing to
    the transverse momentum distribution of a $Z$ boson, which is represented by a blue zigzag line.  The soft
    radiation pattern (dashed line) is associated with the dipoles formed by the outgoing hard gluon (red curly line)
    and the initial state quarks, and is not azimuthally symmetric.  }
\end{figure}

In this work we study in the large-$\bz$ approximation
the transverse momentum distribution of a vector boson, and
in particular, we want to determine whether it is affected by linear renormalons.
Our interest in this process stems from
the fact that, unlike the inclusive Drell-Yan case,
the soft emission pattern in the production of a vector
boson in association with a hard jet
is not azimuthally symmetric, as shown schematically in fig.~\ref{fig:SoftPatternZ}.
Under these circumstances, we can expect that soft gluons
may induce linear renormalon corrections to the
transverse momentum distribution of the vector boson, since they are
not emitted according to an azimuthally symmetric pattern.
We also notice that in ref.~\cite{FerrarioRavasio:2018ubr}, it was found
that certain leptonic observables in top production and decay are affected
by linear renormalons, thus casting doubts on the common assumption that
leptonic observables should be less affected by non-perturbative corrections
with respect to hadronic ones.


\subsection{On the small $p_T$ distribution of vector bosons}

The transverse momentum distribution of vector bosons in the region of small
transverse momenta has been the subject of considerable theoretical activity
since the early days of perturbative QCD~\cite{Dokshitzer:1978yd,Parisi:1979se,Collins:1984kg}
up to the present days~\cite{Becher:2010tm,Bozzi:2010xn,Becher:2011xn,GarciaEchevarria:2011rb,Bizon:2018foh,Becher:2019bnm,HAUTMANN2020135478}.
It has a very peculiar perturbative expansion in QCD, starting with a $\delta^2(p_T)$
at leading order, and receiving singular contributions due to soft and collinear
gluon emissions at higher perturbative
orders. It is dealt with by resumming the singular contributions to all
orders in the perturbative expansion. This turns out to yield (for very large masses)
finite results down
to zero transverse momentum, according to a mechanism first exposed by Parisi and Petronzio
in ref.~\cite{Parisi:1979se}. It was however immediately recognized that some
non-perturbative input is required in these calculations, since the resummation of
soft-collinear gluons must be cut-off at transverse momenta of the order of typical
hadronic scales. Also, by intuitive reasoning, one expects that Fermi motion effects
should smear the $\delta^2(p_T)$ leading order contribution (and the higher order
singular ones) to a transverse size of
the order of a Fermi.\footnote{These effects are also modelled using Transverse Momentum
  Dependent (TMD) parton densities~\cite{Angeles-Martinez:2015sea}.}

Considering for the sake of argument the smearing due to Fermi motion,
let us call $f (\vec{q}_T) \mathd^2 \vec{q}_T$ the primordial transverse momentum distribution of the
quarks in a hadron. If $f (\vec{q}_T)$ is dominated by values of $\vec{q}_T$ of the
order of a hadronic scale, $f$ can be expanded in moments
\begin{equation}
  f (\vec{q}_T) = \delta^2 (\vec{q}_T) + \Lambda^2 (\vec{\partial}_{p_T})^2
  \delta^2 (\vec{q}_T) + \mbox{higher derivatives},
  \label{eqLprimordial}
\end{equation}
where $\Lambda$ is a typical hadronic scale,
corresponding in parameter space to the behaviour
\begin{equation}
  \tilde{f}(\vec{b}) = \int \mathd{}^2p_T e^{i \vec{q}_T\cdot \vec{b}}f (\vec{q}_T)=
  1-\Lambda^2 b^2+\mbox{higher orders in $b$}. \label{eqLprimordialb}
\end{equation}
Notice that linear terms in  $\partial_{p_T}$ (or $b$) are naturally absent (at least as long as one does not
consider spin structures in the distribution), and in fact the behaviour
given in eq.~(\ref{eqLprimordialb}) has been advocated as
early as ref.~\cite{Parisi:1979se}, where the form
$\tilde{f}(b)=\exp(-\Lambda^2 b^2)$ was proposed.
The same form also appears in several subsequent works.\footnote{
  More refined theoretical analysis of these non-perturbative corrections can be found in~\cite{Becher:2013iya,Scimemi:2016ffw}.
  In the context of TMD parton distributions, the absence of linear power corrections can also be seen formally as
  a consequence of the operator product expansion that they obey.
}
When convoluted with perturbative mechanisms giving rise to the
transverse momentum of the vector boson, the form of eq.~(\ref{eqLprimordial})
leads to corrections of order $\Lambda^2/p_T^2$.

The absence of linear corrections in this context has also a rather simple intuitive
explanation. The primordial transverse momentum smearing gives a
transverse kick, of the order of typical hadronic scales,
to the perturbative distribution.
However, it is azimuthally symmetric. Thus, its first-order effects
cancel out, leaving only quadratic corrections.
On the other hand, the non-perturbative
corrections introduced in this context are all that is needed in order to
regularise the ill-behaved perturbative expansion
in the small $p_T$ region. They cannot be claimed to be
the \emph{only} non-perturbative corrections that are relevant to the problem,
especially if we move to regions of phase space where the QCD perturbative
expansion is totally well-behaved.

At variance with the case of the small transverse momentum calculations,
we deal with a process that has a well
behaved perturbative expansion in the strong coupling constant.
We do not make any assumption regarding the origin of non-perturbative
corrections. We compute the full cross section, and look for
renormalon effects in the full result. We can only do this, however, by
considering a simpler theory, i.e. the large $n_f$ limit of QCD, where we can
compute the cross section exactly, at all orders in the perturbative
expansion.

We expect that linear effects may appear only if the soft radiation pattern is
not azimuthally symmetric, and thus we considered a process where
such asymmetric configuration is realized. For example, we did not consider
the process where two incoming partons produce a $Z$ plus a photon with large
transverse momentum. Such process does yield an
azimuthally symmetric pattern for gluon radiation, and thus we do not
expect linear renormalons to arise there.

We also remark that the thrust of a jet does receive linear power corrections
associated with linear renormalons
(see for example eq.~(5.56) in ref.~\cite{Beneke:1998ui}).
Since in the process we consider
the $Z$ is recoiling against a jet, the worry that such linear
corrections may affect the $Z$ transverse momentum
distribution is not without ground.


\subsection{Our calculation}
\begin{figure}[tb]
  \centering
  \includegraphics[width=0.6\linewidth]{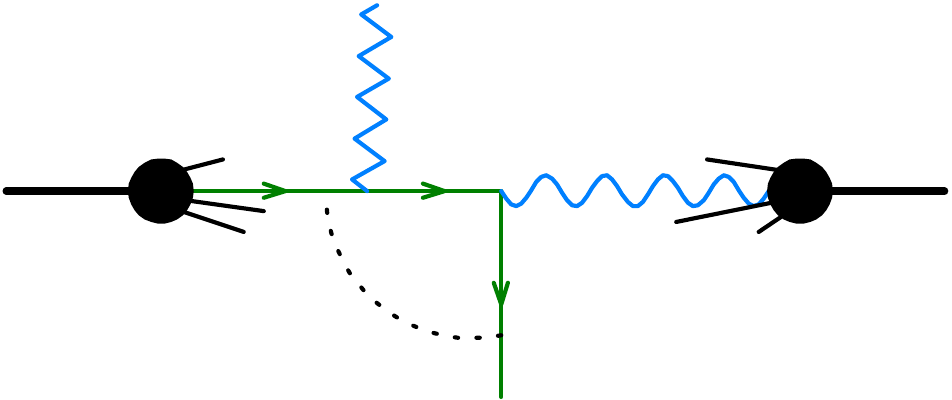}
  \caption{\label{fig:qgZ}  A Born diagram for $Z$ production in photon-quark
    collision. The incoming photon is represented by the blue wavy line.
    This process has an azimuthal asymmetric colour pattern
    of soft emission, and is suitable for testing for the presence of linear
    renormalons in the $Z$ transverse momentum distribution.}
\end{figure}

\begin{figure}[tb]
  \centering
  \begin{displaymath}
    \stackrel{\includegraphics[width=0.9\linewidth]{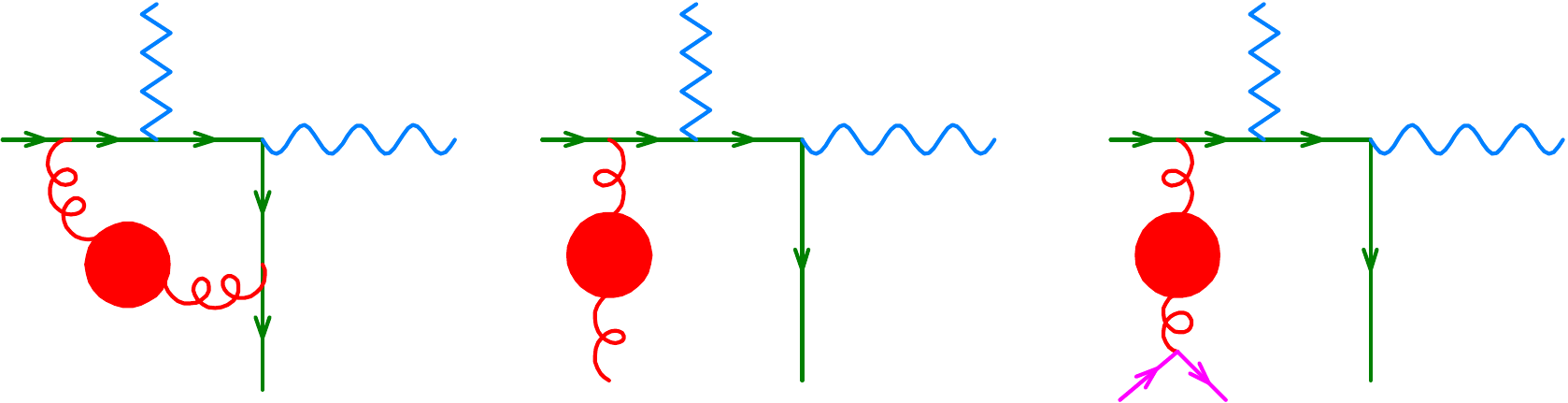}}{\phantom{zzzz}({\rm v})
      \phantom{zzzzzzzzzzzzzzzzzzzzzz}({\rm g})\phantom{zzzzzzzzzzzzzzzzzzzzz}({\rm q\bar{q}})}
  \end{displaymath}
  \caption{\label{fig:Largebz} A sample of the radiative
    corrections that need to be included in order to compute the leading
    large $\bz$ corrections to the process of fig.~\ref{fig:qgZ}.}
\end{figure}

A calculation of the large-$\bz$ corrections to the process depicted in fig.~\ref{fig:SoftPatternZ}
is too demanding, and in fact, at the moment, no large-$\bz$ corrections to processes
already involving a gluon emission or exchange has been ever carried out. Such calculation
would inevitably involve dressed gluon propagators joining into a three-gluon vertex,
also including a vertex correction formed by a quark triangle graph (such correction is in fact
of order $g_s(g_s^2 n_f)$). Thus, we will instead
compute the process depicted in fig.~\ref{fig:qgZ},
i.e. the production of a $Z$ boson in photon-quark collisions.
More specifically, we assume that the incoming photon couples to a single quark flavour
(that we will call for definiteness a $d$ quark)
but there is a large number $\nf$ of light quark flavours that we denote with the symbol $q$.
We can thus compute this process
in the large $\nf$ limit, and turn to the  large $\bz$ limit in the usual way, i.e.
by replacing the $b_0$ in the Abelian theory with the full QCD $b_0$. In this process
(as in the realistic QCD one)
the soft emission pattern is not azimuthally
symmetric, and we can investigate whether linear IR renormalons are present
in the $Z$ transverse momentum distribution due to recoil against soft emissions.
The radiative corrections that one needs
to compute are represented schematically in fig.~\ref{fig:Largebz}.
The solid blob insertion in the gluon propagator represents the inclusion of all
corrections given by a fermion loop, as represented
by the recursive graphic equation
\begin{equation}\label{eq:blobeq}
  \raisebox{-0.5cm}{\includegraphics[width=0.7\linewidth]{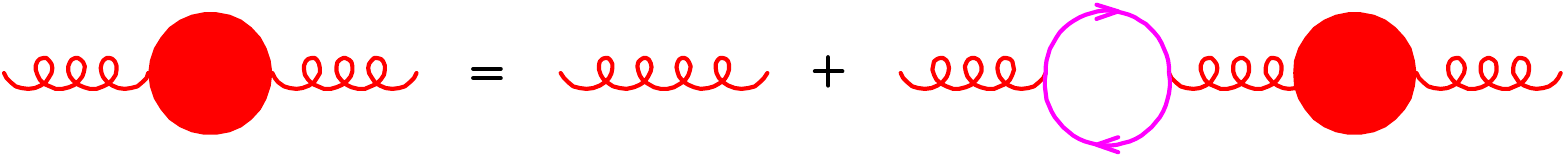}}\,.
\end{equation}
The inclusion of all corrections embodied in eq.~(\ref{eq:blobeq}) amounts to
considering $\as \nf$ to be of order 1. Because of this reason, the gluon
splitting into fermion pairs must also be included, since, when squared,
it also gives rise to a factor $\as \nf$.
We also notice that we do not need to worry about the interference of the fermions arising from
gluon splitting with the fermion coming from the initial hadron, since this term is suppressed
by a factor of $\nf$.
We have developed a parton-level generator for the computation of this process
that can be used for the computation of
any IR finite differential distribution,
provided a cut in the $Z$ transverse momentum is included.

The rest of the paper is organized as follows. In Section~\ref{sec:method} we describe
the procedure
that we adopt in order to perform the calculation. In essence, the result is expressed in terms
of a corresponding next-to-leading correction to the Born process in a fictitious theory where
the gluon has a finite mass, plus a ``remainder'' correction
that accounts for the difference of the calculation in the massive case with respect to
the one performed in the full theory~\cite{Nason:1995np,Dokshitzer:1997iz,Dokshitzer:1998pt}.
This difference does not
contribute to quantities that do not depend upon the kinematics of coloured
objects in the final state, and thus does not contribute to the transverse
momentum distribution of the $Z$ boson.
The procedure for the removal of the
initial state collinear singularities is discussed here.

In Section~\ref{sec:CalcDetails} we give further
details on the setup of the calculation. In particular, it is shown that the real cross section must
be partitioned into three terms, each of them appropriate to one of the singular regions
of the real cross section, and each term (requiring an appropriate integration importance sampling)
is computed independently. Finally, in section~\ref{sec:Results} we present the results of our calculation.
In sec.~\ref{sec:Conclusions} we give our conclusions.

\section{The method}\label{sec:method}
We assume we are calculating a cross section with a given set of cuts, that we represent with a function of the
kinematic configuration $\Theta(\Phi)$, that takes the value 1 if the cuts are satisfied, and zero otherwise. The cross section for our process (borrowing from the notation of ref.~\cite{Frixione:2007vw}) is given by
\begin{eqnarray}
  \sigma&=&\int \mathd \PhiB \left(\sB(\PhiB)+ \sV(\PhiB)\right) \,\Theta(\PhiB) \nonumber \\
        &+& \int \mathd \Phiplus \splus(\Phiplus) \,\Theta(\Phiplus) \nonumber \\
        &+& \int \mathd \Phiminus^{g} \sminus^{g}(\Phiminus^{g}) \,\Theta(\Phiminus^{g})
            + \int \mathd \Phiminus^{q{\bar q}} \sminus^{q{\bar q}}(\Phiminus^{q{\bar q}}) \,\Theta(\Phiminus^{q{\bar q}})
            \nonumber \\
        &+& \int \mathd \Phig \sg(\Phig) \,\Theta(\Phig) \nonumber \\
        &+&\int \mathd \Phiqqb \sqqb(\Phiqqb) \Theta(\Phiqqb)\,,
\end{eqnarray}
where $\sB$ represents the Born term for the process $d\gamma \to d Z$ (see fig.~\ref{fig:qgZ}),
$\sV$ refers to the virtual correction to the  $d\gamma \to d Z$ process, $\sg$
refers to the process $d\gamma \to d Z g$, and  $\sqqb$ to the process  $d\gamma \to d Z \qi \qbi $
(see the diagrams labelled as $({\rm v})$, $({\rm g})$ and $({\rm q\bar{q}})$ in fig.~\ref{fig:Largebz}).
The $\splus$ term is the subtraction of the collinear singularities arising in both  $\sg$
and  $\sqqb$ when the gluon or the $\qi\qbi$ pair becomes collinear with the incoming quark. Both these
collinear configurations are associated with the same underlying Born $d\gamma \to d Z$, and thus can
be represented by a single term.
The $\sminus^{g}$ and  $\sminus^{\qi\qbi}$ terms represent the collinear counterterms associated
with the collinear singularity due to the
photon splitting into a $d {\bar d}$ pair in the $\sg$ and $\sqqb$ contributions respectively.
We have defined
\begin{eqnarray}
  \mathd \PhiB &=& \mathd \xplus \mathd \xminus \mathd \phiB, \label{eq:bfphiBdef} \\
  \mathd \phiB &=& \frac{\mathd^3 \veckz}{2\kz^0(2\pi)^3} \frac{\mathd^3\veckd}{2\kd^0(2\pi)^3}
                   (2\pi)^4 \delta^4(\pplus \xplus +\pminus \xminus - \kd - \kz)\,, \label{eq:phiBdef}
\end{eqnarray}
where $\pplus$, $\pminus$ are the momenta of the incoming positive and negative rapidity hadrons,
$\xplus$, $\xminus$ are the momentum fractions of the incoming light quark and photon,
and $\kz$ and $\kd$ are the momenta of final state $Z$ and light quark $d$.
Furthermore
\begin{equation}
  B(\PhiB)=f_d(\xplus) f_\gamma(\xminus) {\cal B} (\pplus \xplus,\pminus \xminus, \kd,\kz)\,, \label{eq:Bdef}
\end{equation}
where ${\cal B}$ is the squared amplitude for the process divided by the flux factor, and $f_d$ ($f_\gamma$)
are the incoming quark (photon) distribution functions.   
The definitions of $\mathd \Phig$ and $\mathd \Phiqqb$, as well as
$\mathd \phig$, $\mathd \phiqqb$ and $R_g$, $R_{q\bar{q}}$ are fully analogous to eqs.~(\ref{eq:bfphiBdef}), (\ref{eq:phiBdef}) and (\ref{eq:Bdef}).

\begin{figure}[tb]
  \centering
  \begin{displaymath}
    \stackrel{\includegraphics[width=0.56\linewidth]{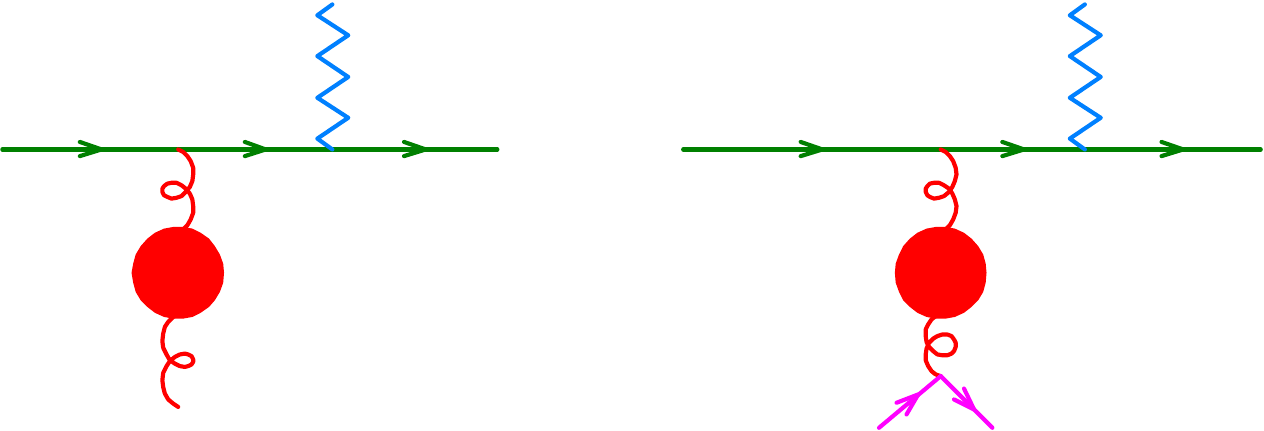}}{\phantom{zzzzzzzzzz}\Bqqb%
      \phantom{zzzzzzzzzzzzzzzzzzzzzzz}\Rqqb\phantom{zzzzzzzzzzzzzz}}
  \end{displaymath}
  \caption{\label{fig:qqbarprocs} Representative graphs of the $d{\bar d}$ initiated processes arising as
    subprocesses in the collinear limit for the incoming photon splitting, contributing respectively
    to the $\Bqqb$ and $\Rqqb$ squared amplitudes.}
\end{figure}

For  $\Phiplus$, $\Phiminus$,  and $\splus$, $\sminus$ we have
\begin{eqnarray}
  \mathd \Phiplus &=&  \mathd \xplus \mathd \xminus \frac{\mathd z}{z} \mathd \phiB,\\
  \mathd \Phiminus^{g} &=&  \mathd \xplus \mathd \xminus \frac{\mathd z}{z} \mathd \phiD^{g},\\
  \mathd \Phiminus^{q{\bar q}} &=&  \mathd \xplus \mathd \xminus \frac{\mathd z}{z} \mathd \phiD^{q{\bar q}}
  \label{eq:bfphiplus}
\end{eqnarray}
and
\begin{eqnarray}
  \splus(\Phiplus)&=&f_d\left(\frac{\xplus}{z}\right) f_\gamma(\xminus)\, C_{dd}(z)\, {\cal B} (\pplus \xplus,\pminus \xminus, \kq,\kz)\,, \label{eq:splusdef} \\
  \sminus^{g}(\Phiminus^{g})&=& f_d(\xplus) f_\gamma\left(\frac{\xminus}{z}\right)\, C_{d\gamma}(z)\, \Bqqb (\pplus \xplus,\pminus \xminus, \kz,\kg)\,. \label{eq:sminusqgdef} \\
  \sminus^{q{\bar q}}(\Phiminus^{q{\bar q}})&=& f_d(\xplus) f_\gamma\left(\frac{\xminus}{z}\right)\, C_{d\gamma}(z)\, \Rqqb (\pplus \xplus,\pminus \xminus, \kz,\kq,\kbarq)\,. \label{eq:sminusqqbdef}
\end{eqnarray}
Thus, the two $\sminus$ collinear subtractions 
correspond to diagrams (g) and  ($\rm{q\bar{q}}$) of fig.~\ref{fig:Largebz}, when the outgoing quark connected to the incoming quark
line becomes collinear to the incoming photon. The $\Bqqb$ and $\Rqqb$ amplitudes correspond
to processes initiated by a $q{\bar q}$ collision with a $Zg$ and a $Zq{\bar q}$ final state respectively,
represented in fig.~\ref{fig:qqbarprocs},
and $\phiD^g$, $\phiD^{q\bar{q}}$ are the corresponding
phase spaces. The $C_{dd}$ and $C_{d\gamma}$ functions are the universal collinear divergent factors
for the $d\to d+X$ and $\gamma\to d + X$ splittings.

We remark that no interference arises in our approximation from the final state down quark connected
to the incoming quark line and the final state quarks arising from gluon splitting. In fact, such
interference term would be down by a factor of $\nf$.

\begin{figure}[tb]
  \centering
  \includegraphics[width=0.6\linewidth]{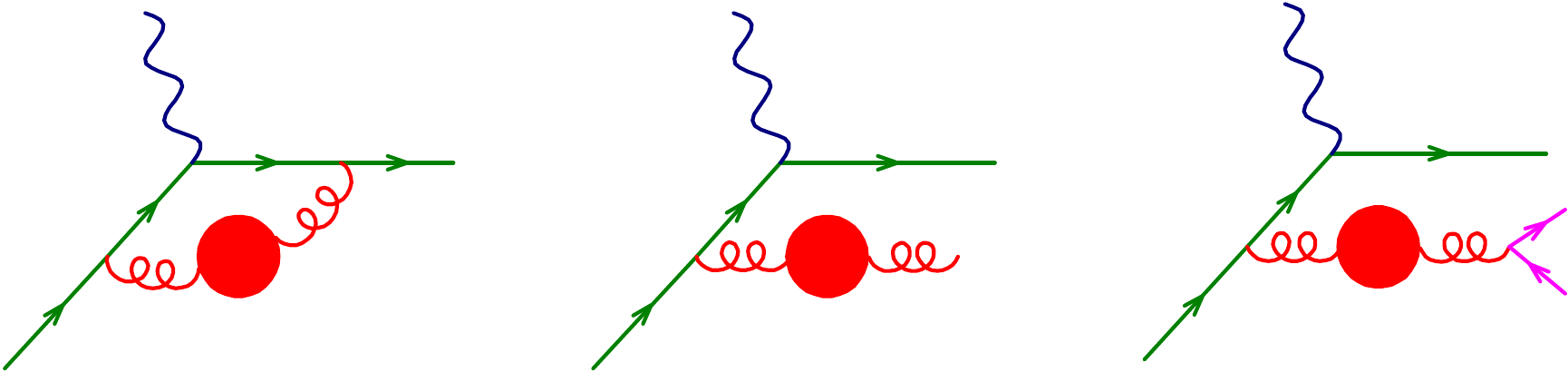}
  \caption{\label{fig:DIS} Representative graphs for the calculation of $F_2$ in the large $\nf$ limit.}
\end{figure}

The diagrams of fig.~\ref{fig:Largebz} are affected
by both ultraviolet and infrared divergences, that we regulate using dimensional regularization.
It turns out that ultraviolet divergences cancel, since they are all associated with vertex and propagator
corrections in an Abelian model. For the collinear subtraction $\sminus$, $C_{d\gamma}$
is given by
\begin{equation}
  C_{d\gamma}(z)=\frac{\alpha \cF^2}{2\pi} \frac{1}{\epsilon} (z^2+(1-z)^2)\,,
\end{equation}
where $\alpha$ is the electromagnetic coupling and $\cF$ is the down-quark charge. This must be
accompanied by the rescaling $\muF^2 \to \muF^2 \exp(-\gamma_{\rm E})/(4\pi)$
according to the \MSB{} subtraction prescription. The \MSB{} subtraction term for the $\splus$
collinear singularity is much more complicated, since it must include all corrections of order $\as \nf$,
i.e. all vacuum polarization insertions in the emitted gluon, and the gluon splitting into a $q{\bar q}$ pair.
Here we avoid this problem (following ref~\cite{Beneke:1995pq}), and
perform our collinear subtraction using the so-called DIS scheme, that is defined by requiring that the structure
function $F_2$ has the expression
\begin{equation}
F_2(x,Q^2)= x \sum_i q_i(x,Q^2) C_i^2,
\end{equation}
(where $C_i$ are the electric charges of the species $i$, and $i$ runs over all quarks and antiquarks)
to all orders in perturbation theory. In this way we simply need to compute $F_2(x,Q^2)$ in the same approximation
that we have used for our process, i.e. by including all fermion loop insertions in the gluon propagator, and then
express our cross section in terms of the $q_i$ in the DIS scheme. The diagrams involved are shown in fig.~(\ref{fig:DIS}).

When translating the DIS scheme cross section into the \MSB{} scheme no new
linear renormalons arise.  This is a consequence of the fact that
$F_2$ obeys an operator product expansion where power corrections are
controlled by the twist of the operator, and the dominant power
corrections, corresponding to twist 4, are quadratic.

The computation of the large $\nf$ cross section can be technically
performed by computing a similar process with a finite gluon mass
$\lambda$.  The result is identical to the one in eq.~(3.22) of
Ref.~\cite{FerrarioRavasio:2018ubr}, with the inclusion of the
collinear remnants, that must be included if the process involves
initial-state singularities.  More specifically, we have
\begin{equation}\label{eq:mainrenormform}
  \sigma=\sigma_B-\frac{1}{\bz\as}\int_0^\infty \frac{\mathd \lambda}{\pi} \frac{\mathd T(\lambda)}{\mathd \lambda}
  \arctan \frac{\pi \bz \as}{1+\bz \as \log\frac{\lambda^2}{\muC^2}}
\end{equation}
where
\begin{equation}
  \as = \as(\mu),\quad\quad \muC=\mu e^\frac{C}{2}, C=\frac{5}{3}, \bz=-\frac{\TF}{2\pi},
\end{equation}
\begin{eqnarray}
\sigma_B&=& \int  \mathd \PhiB \, \sB(\PhiB) \,\Theta(\PhiB)\,, \\
  T(\lambda) &=& T_V(\lambda) + T_\oplus(\lambda)+T_\ominus(\lambda)+T_\ominus^\Delta(\lambda) + T_R(\lambda) + T_R^\Delta(\lambda)\,, 
 \label{eq:Tlambda} \\
  T_V(\lambda) &=& \int  \mathd \PhiB \, \sV^{(\lambda)}(\PhiB)  \,\Theta(\PhiB)\,, \\
  T_\oplus(\lambda) &=& \int \mathd \Phiplus
                        f_d\left(\frac{\xplus}{z}\right) f_\gamma(\xminus)\, C^{(\lambda)}_{dd}(z)\,
                        {\cal B} (\pplus \xplus,\pminus \xminus, \kd,\kz) \,\Theta(\Phiplus)\,,  \\
  T_\ominus(\lambda) &=& \int \mathd \Phiminus^{g^*} f_d(\xplus) f_\gamma\left(\frac{\xminus}{z}\right)\, C_{d\gamma}(z)\,
  \Bqqbstar (\pplus \xplus,\pminus \xminus, \kz,\kgstar)\,\Theta(\Phiminus^{g^*})\,,  \label{eq:Tominus}\\
  \label{eq:tdeltaminus}  T_\ominus^\Delta(\lambda) &=& \int \mathd \Phiminus^{q\bar{q}} f_d(\xplus) f_\gamma\left(\frac{\xminus}{z}\right)\, C_{d\gamma}(z)\,
                                {\Rqqb} (\pplus \xplus,\pminus \xminus, \kz,k_q,k_{\bar{q}})\nonumber \\
        &\times& \delta\left(\lambda^2-\kqqb^2\right)
                 \left[\Theta(\Phiminus^{q\bar{q}})- \Theta(\Phiminus^{g^*})\right] 
\end{eqnarray}
\begin{eqnarray}
  T_R(\lambda) &=&
   \int \mathd \Phigstar \, \sgstar(\Phigstar)  \,\Theta(\Phigstar)
  \,, \\
\label{eq:trdelta}  T_R^\Delta(\lambda) &=& \frac{3\pi}{\as \TF} \lambda^2
                      \int \mathd \Phiqqb\, \delta(\lambda^2-k_{q\bar{q}}^2)\,\sqqb(\Phi_{q{\bar q}})
                      \left[\Theta(\Phiqqb)-\Theta(\Phigstar) \right]\,.
\end{eqnarray}
With $g^*$ we denote a gluon with mass $\lambda$, and with the superscript $(\lambda)$ applied to
previously defined objects we denote the same objects computed for
a massive gluon with mass $\lambda$. Thus, for example, $\sV^{(\lambda)}$ are the virtual corrections depicted
schematically in fig.~\ref{fig:Largebz} and labelled with (v), where the gluon propagator with the
red blob is replaced by a massive gluon propagator with gluon mass $\lambda$, $\Phigstar$ denotes the
phase space for the process $d\gamma \to Z d g^* $, and so on.
In eqs.~(\ref{eq:tdeltaminus}) and~(\ref{eq:trdelta}) we have defined $\kqqb=k_\qi+k_\qbi$.

The fact that the large $\nf$ calculation can be given in terms of cross sections involving a massive gluon
is well known. The appendix of ref.~\cite{FerrarioRavasio:2018ubr} summarises all the technical steps that lead to
this equivalence, but this is not the reference where this equivalence has been found first.
It has been used in ref.~\cite{Beneke:1995pq} in order to compute the Drell-Yan and DIS process in the
large $\nf$ limit. The need of the $\Delta$ terms has been first pointed out in ref.~\cite{Nason:1995np},
and is the only correction that is needed to connect the massive gluon calculation to the realistic
case, where the virtual gluon decays into a $q{\bar q}$ pair.

For the purpose of the present work, the $\Delta$ terms do not contribute. In fact, our cuts only depend
upon the kinematics of the $Z$ boson, and thus are insensitive to whether we replace the $q{\bar q}$ pair
arising from gluon splitting with an undecayed massive gluon with the mass equal to the virtuality of the pair.
Thus, the differences of theta
functions in the square brackets are always zero in our case.

The $C^{(\lambda)}_{dd}(z)$ term was computed in eq.~(3.10) of ref.~\cite{Beneke:1995pq}, and in terms
of the quantities defined there we have
\begin{equation}
  C^{(\lambda)}_{dd}(z)=-f^{\rm (1), real} \left(z,\frac{\lambda^2}{\mu^2}\right)
  -\delta(1-z) \,f^{\rm (1), virt} \left(\frac{\lambda^2}{\mu^2}\right)\,.
\end{equation}
The $z$ variable in the $f^{\rm (1), real}$ term is limited to the range
\begin{equation}
  z<z_{\rm max}\equiv \left(1+\frac{\lambda^2}{\mu^2}\right)^{-1}.
\end{equation}
As a consequence of the Adler sum rule, we have
\begin{equation}\label{eq:Adler}
  \int_0^1  C^{(\lambda)}_{dd}(z)\, \mathd z =
  - \int_0^{z_{\rm max}} f^{\rm (1), real}
  \left(z,\frac{\lambda^2}{\mu^2}\right)\, \mathd z  -f^{\rm (1), virt} \left(\frac{\lambda^2}{\mu^2}\right)
  =0\,.
\end{equation}

Formula (\ref{eq:mainrenormform}) relates the massive calculation with the full leading $\nf$ perturbative
expansion. We remark that the integrand in (\ref{eq:mainrenormform}) represents formally
a power expansion in $b_0\as$ with factorially growing coefficients.
In fact, it turns out that $T(\lambda)$ is finite as $\lambda \to 0$, and it has an expansion in $\lambda$
of the form
\begin{equation}
  T(\lambda) = T(0) + T'(0) \lambda + {\cal O}\left(\lambda^2\right),
  \label{eq:Tslope}
\end{equation}
with the possible presence of logarithmic enhancements in the quadratic remainder.
If $T'(0)\neq 0$ then the perturbative expansion of the result has a linear renormalon. Its structure is examined in detail in appendix~\ref{sec:AtanForm}.

The computation of the collinear subtraction arising from the collinear splitting of the incoming photon
is a standard NLO subtraction, and can be handled by standard means, as those already coded in
the \POWHEGBOX{} package~\cite{Alioli:2010xd}. It is useful, however, to clarify from the beginning that
we do not expect any linear renormalon corrections to the $Z$ transverse-momentum distribution
from this collinear region.
In fact, as stated earlier, the $T_\ominus^\Delta(\lambda)$ term does not contribute, and if the final state
quark is collinear to the incoming photon, the transverse momentum of the $Z$ must be balanced by
the massive gluon. Since we are only considering distributions where the $Z$ has a sizable transverse
momentum, the massive gluon will also have a large transverse momentum, and under these conditions
the mass correction to the process is quadratic in $\lambda$.

\section{Details of the calculation}\label{sec:CalcDetails}
\begin{figure}[tb]
  \centering
  \includegraphics[width=0.7\linewidth]{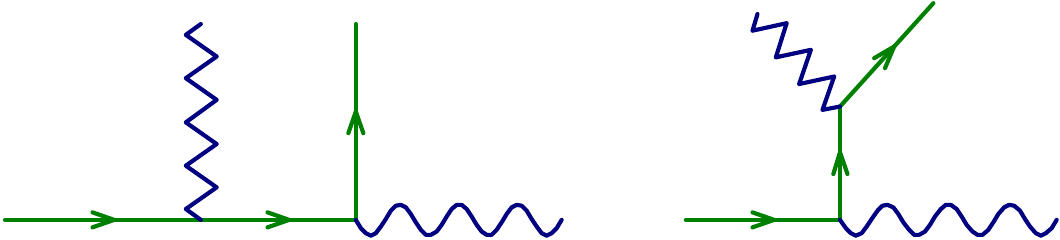}
  \caption{\label{fig:Bornqg} The amplitudes for the Born contribution in the $q\gamma\to Z q$
    process. The zigzag line represents the $Z$ boson, while the wavy line represents the
    incoming photon. }
\end{figure}

As illustrated in the previous section, the computation of the IR renormalon contribution
requires us to perform the calculation in the context of an Abelian theory with a massive gluon.
The Born diagrams for our process are represented in fig.~\ref{fig:Bornqg}.
For simplicity, we have taken the $Z$ boson to be vectorially coupled and stable.
The Born and Virtual contributions are computed analytically at fixed external momenta.
The computation was performed using the symbolic manipulation program MAXIMA~\cite{maxima},
using a package for the computation of Feynman Diagrams developed by one of the authors~\cite{hepmaxima}. One-loop scalar integrals have been evaluated using the {\tt COLLIER} library~\cite{{Denner:2016kdg}}.
The virtual corrections are obtained by attaching a (massive) gluon propagator to two points along the fermion line
in all possible ways, avoiding however to attach both ends to the
same external line.
Virtual corrections to the external lines are instead dealt with according to the usual LSZ prescription,
i.e. one multiplies the Born diagram by the wave function renormalization correction. The virtual
contribution to the cross section is computed by multiplying the sum of the virtual amplitudes
with the conjugate of the sum of Born amplitudes, and by taking twice the real part of the result.
The virtual corrections are infrared finite, since the gluon mass regulates infrared divergences.
However, each individual contribution is affected by ultraviolet divergences, that
we regulate using standard dimensional regularization, so that gauge invariance is preserved
at every step of the calculation. The sum of all contributions is ultraviolet finite,
and thus one can take its limit to 4 dimensions.

Upon integration over the external momenta, the Born and Virtual contributions are finite as
long as we require a lower bound on the $Z$ transverse momentum. The numerical treatment
of this stage of the calculation will be detailed further on.

\begin{figure}[tb]
  \centering
  \includegraphics[width=0.6\linewidth]{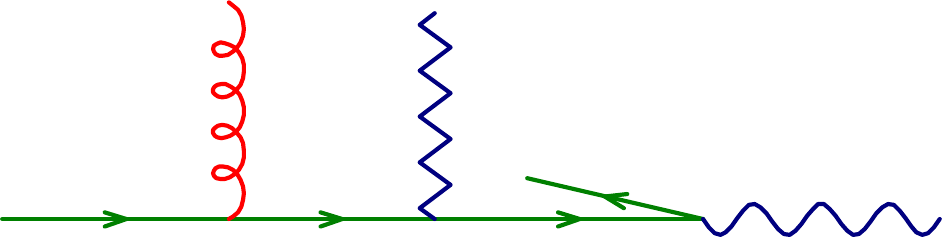}
  \caption{\label{fig:RealCollSing} Configuration that leads to collinear singularities
    in the $q\gamma\to Z q g$ process.}
\end{figure}

The real corrections are obtained by attaching one external massive gluon in all possible ways
to the Born graph. Upon phase space integration, real corrections are not infrared finite even if one
requires a finite transverse momentum of the $Z$-boson, due to the collinear singularities
arising from the photon splitting into two massless quarks, as shown
in fig.~\ref{fig:RealCollSing}.
This kind of singularities should also be regulated in dimensional regularization, and an \MSB{}
subtraction should be performed as required by the factorization formalism. In fact, this procedure
is automatically implemented in the \POWHEGBOX{}, and does not require any further work on our part.

\begin{figure}[tb]
  \centering
  \includegraphics[width=\linewidth]{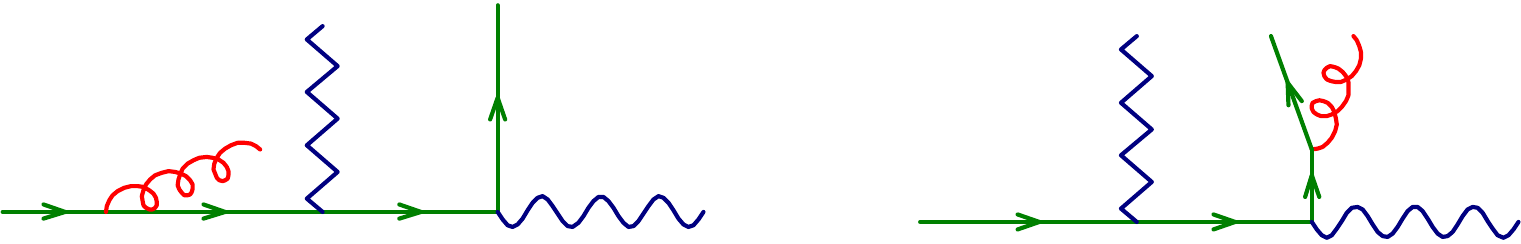}
  \caption{\label{fig:RealSing23} Singularities in the $\lambda \to 0$ limit.
    The diagram on the left contributes a  collinear initial state
    singularity, and a soft singularity. The diagram on the right contributes a final state
    collinear singularity, and a soft singularity.}
\end{figure}

Singularities associated with soft or collinear gluon emissions are instead regulated by the gluon mass.
These are illustrated in fig.~\ref{fig:RealSing23},
corresponding to an initial and a final-state collinear singularity, associated with a soft one.
Real contributions thus behave as  $\log^2\lambda$ for small $\lambda$.
The same $\log^2\lambda$ term, but with opposite sign, comes from the virtual corrections, and
cancels the real one.
After the cancellation of the double logarithmic terms,
the only remaining singularity is a single logarithm of $\lambda$,
that cancels against an opposite contribution in the collinear counterterm (see eq.~(\ref{eq:Tominus})).
At this point we get a result that has a finite limit as $\lambda\to 0$: the cross
section goes to a constant and our task is to determine whether this constant is
approached with a slope linear in $\lambda$.

It is clear that, in view of the massive cancellations involved, in
order to get a convincing numerical evidence of the small $\lambda$
behaviour of the cross section, the numerical integration should be performed
with an appropriate importance sampling near the regions that are
singular in the $\lambda \to 0$ limit. Furthermore, a direct
calculation in the massless case, i.e. at $\lambda=0$, is also
necessary, since it gives us a point with negligible error, that would
be difficult to obtain with small values of $\lambda$. In order to perform the calculation,
the real contribution is separated into three terms:
\begin{equation}
  R=R^{(1)}+R^{(2)}+R^{(3)}
\end{equation}
\begin{eqnarray}
    R^{(1)} &=& \frac{\frac{1}{p_{\rm \scriptscriptstyle T,d}^2} }{\frac{1}{p_{\rm \scriptscriptstyle T,d}^2}+\frac{1}{m_{\rm \scriptscriptstyle T,d}^2}+\frac{(E_d+E_g)^2}{E_d E_g m_{d,g}^2}} \,R, \\
\label{eq:R-region-1}
  R^{(2)} &=& \frac{\frac{1}{m_{\rm \scriptscriptstyle T,g}^2}}{\frac{1}{p_{\rm \scriptscriptstyle T,d}^2}+\frac{1}{m_{\rm \scriptscriptstyle T,d}^2}+\frac{(E_d+E_g)^2}{E_d E_g m_{d,g}^2}} \,R,\\
  R^{(3)} &=& \frac{\frac{(E_d+E_g)^2}{E_d E_g m_{d,g}^2}}{\frac{1}{p_{\rm \scriptscriptstyle T,d}^2}+\frac{1}{m_{\rm \scriptscriptstyle T,d}^2}+\frac{(E_d+E_g)^2}{E_d E_g m_{d,g}^2}} \,R,
\end{eqnarray}
where $p_{\rm \scriptscriptstyle T,d}$ is the transverse momentum of the final-state quark, and $m_{\rm \scriptscriptstyle T,g}$ is the transverse mass
of the final-state gluon, defined as
\begin{equation}
  m_{\rm \scriptscriptstyle T,g}^2=p_{\rm \scriptscriptstyle T,g}^2+\lambda^2,
\end{equation}
$m_{dg}$ is the mass
of the quark-gluon final-state system, and $E_d$, $E_g$ are the energies of the quark and gluon in the partonic center of mass. Notice that the expression
\begin{equation}
  \frac{E_dE_g}{(E_d+E_g)^2} m_{d,g}^2\approx \frac{E_d^2E_g^2}{(E_d+E_g)^2}(1-\cos \theta_{d g})
\end{equation}
is of the order of the transverse mass of the gluon relative to the final state quark.
The three superscripts $(1)$, $(2)$ and $(3)$ label the three regions where each of the three contributions
is singular, i.e. region (1) is singular when the final state quark is collinear to the incoming photon;
region (2) when the gluon is
collinear to the incoming quark, and region (3) when the final state gluon is collinear to the outgoing quark.

The real contributions for regions (2) and (3) are evaluated
independently, and with a different parametrisation of the phase
space. In case of region (2), the phase space is factorized as the
product of the two body phase space formed by the final state gluon
recoiling against the quark-$Z$ system, and the two body phase space
for the quark-$Z$ system itself.  In the case of region (3), the phase
space is factorized in terms of the two body phase space for the
system comprising the $Z$, and the quark-gluon system recoiling
against it, with the quark-gluon system itself parametrised as a two
body phase space.  With these parametrisations it is easy to perform
importance sampling integration near the singular configurations.  We
notice that both the real and virtual contributions, as well as the
Born terms, are singular when the transverse momentum of the $Z$
vanishes. However, since we are interested in the transverse momentum
distribution of the $Z$ for transverse momenta comparable to the $Z$
mass, when computing the integral we can suppress this region with a
factor proportional to the $Z$ transverse momentum. We choose
\begin{equation}
F_{\rm supp}=\frac{\ptz^4}{\ptz^4+\ptcut^4},
\end{equation}
where $\ptcut$ is a parameter close to the transverse
momentum cut that we would like to apply to our cross section.  The
adaptive Monte Carlo integration is performed in a standard way,
multiplying the Real, Born, Virtual and collinear subtraction
contributions by this suppression factor, in order to obtain a
convergent result. This factor is divided out when computing the cross
section with cuts, in order to obtain the correct and finite
result. In fact, since we always require a transverse momentum of the
$Z$ larger than a given cut, $F_{\rm supp}$ never vanishes in the
region of interest.
\subsection{Region $(1)$}
The contribution of region (1) can be evaluated using directly the \POWHEGBOX{}
framework, taking the real process $d\gamma \to Z d g$, and the Born process
$d \bar{d} \to Z g$. The real cross section is taken equal to $R^{(1)}$. The Born
contribution itself is absent, since we assume that our incoming hadrons have only
a down quark (for the positive rapidity incoming hadron), and a photon (for the negative
rapidity one) content. The Born subprocess $d \bar{d} \to Z g$, however, enters in the collinear subtraction, that
is automatically performed by the  \POWHEGBOX{} to implement the factorization
of collinear singularities, as well as in the collinear remnant, that is also automatically
computed by the  \POWHEGBOX{} framework.

The region of phase space dominated by soft and collinear gluons is expected to be highly suppressed in region $(1)$, to such an extent that we do not expect any linear renormalons here. In fact, the collinear and soft-collinear
region associated with an ISR gluon emitted by the quark is suppressed by a $m_{\rm \scriptscriptstyle T,g}^2$ factor (see eq.~(\ref{eq:R-region-1})). 
In this singular limit, for small $\lambda$, $R^{(1)}$ behaves like
\begin{equation}
  \frac{d \theta_g}{\theta_g} \frac{d E_g}{E_g} m_{\rm \scriptscriptstyle T,g}^2 \sim \frac{d\theta_g}{\theta_g}\frac{d E_g}{E_g}
  \theta_g^2 E_g^2\sim  dE_g^2 \,d \theta_g^2,
\end{equation}
where $E_g$ is the gluon energy, $\theta_g$ its angle with respect to the incoming (or outgoing) quark.  The gluon mass $\lambda$ provides a lower cutoff on the $E_g$ and $\theta_g$ integrations, so we see that this term cannot develop a linear sensitivity to $\lambda$. 
\subsection{Computation of the $T(\lambda)$ functions}
For every observable that we consider, we compute the associated $T(\lambda)$ function.
The calculation is performed as follows:
\begin{itemize}
\item
  The $\lambda=0$ contribution is computed by implementing our process in the \POWHEGBOX{} framework. The contributions
  of regions 2 and 3 are computed together. They have singular regions associated with initial state radiation of a gluon
  from the (positive rapidity) incoming quark, and final state radiation of a gluon from the final state quark. The collinear
  singularity associated with a vanishing transverse momentum of the outgoing quark is absent in regions 2 and 3. Care must be
  taken to perform the subtraction of the initial state collinear singularity, that, for consistency, has to be performed
  in the DIS scheme. This affects the collinear remnant, and we suitably modified the \POWHEGBOX{} code in order to
  comply with this request. The factorization scale, corresponding to the scale $Q$ of the DIS scheme subtraction,
  is taken equal to the $Z$ mass. The virtual corrections to the process $d\gamma \to Z d$ enter this calculation, and
  are computed directly for $\lambda=0$ using dimensional regularization. The cancellation of the associated soft
  and collinear divergences with those arising from the real contributions is already implemented in the \POWHEGBOX{}
  in a general way, and requires no further action.
\item  The calculation for $\lambda\neq 0$ for the regions 2 and 3 cannot be simply implemented in the \POWHEGBOX{}
  framework, that was not designed to handle singularities regulated by a mass. 
  They were thus implemented in a dedicated Fortran
  code. In particular, for the real contribution care was taken to adopt phase space parametrisations that are suitable
  to handle each region with adequate importance sampling. The virtual contribution was also evaluated independently.
  Rather than computing separately the real and virtual contributions of the collinear subtraction, we implemented
  a local cancellation of the associated soft divergence by making use of relation~(\ref{eq:Adler}).
\item
  The contribution from region 1,
  for either $\lambda=0$ or $\lambda\neq 0$, is also implemented in the
  \POWHEGBOX{}.  In this case the only singular region present is
  the one associated with the collinear splitting of the
    initial state photon into a $d\bar{d}$ pair. The
    underlying Born for this singularity is given by the
    $d\bar{d}\to Z g$ process, where the gluon has mass $\lambda$,
    different from the case of regions~2 and~3. The collinear singularity is
    treated in the \MSB{} scheme in this case, and the collinear remnant
    is automatically provided by the \POWHEGBOX{}.
\end{itemize}

\section{Results}\label{sec:Results}
As our benchmark set-up, we have taken two colliding particles with center-of-mass (CM) energy of $300\;$GeV.
The positive rapidity incoming particle (labelled as $(1)$) has a parton density consisting only of down quarks,
while the negative rapidity particle (labelled as $(2)$) has a parton density consisting only of photons, distributed as
\begin{equation}
f^{(1)}_d(x)=f^{(2)}_\gamma(x)=\frac{(1-x)^3}{x}.
\end{equation}
This totally arbitrary choice is only dictated by simplicity, and is adequate for our purposes.
We compute the cross section for the production of a stable vector boson $Z$ of mass $M_Z=91.188{\,}\text{GeV}$,
that is only vectorially coupled. The $Z$ and $\gamma$ couplings are both given by $g_{Z/\gamma}^2=4\pi$,
and the down quark is taken to have charge $-1/3$.\footnote{The actual values of the couplings are irrelevant for our
  conclusions, and are only presented to give
  a well-defined meaning to our
  numerical results.}
The Born diagrams have been computed supplying the correct colour factor (that is 1), and
in the calculation of the virtual and real corrections we have included the appropriate QCD colour factor $C_F$.
We have chosen the factorization scale $\muF=M_Z$. The renormalization scale
choice does not affect $T(\lambda)$.

\begin{figure}[tb]
  \centering
    \includegraphics[width=0.48\textwidth]{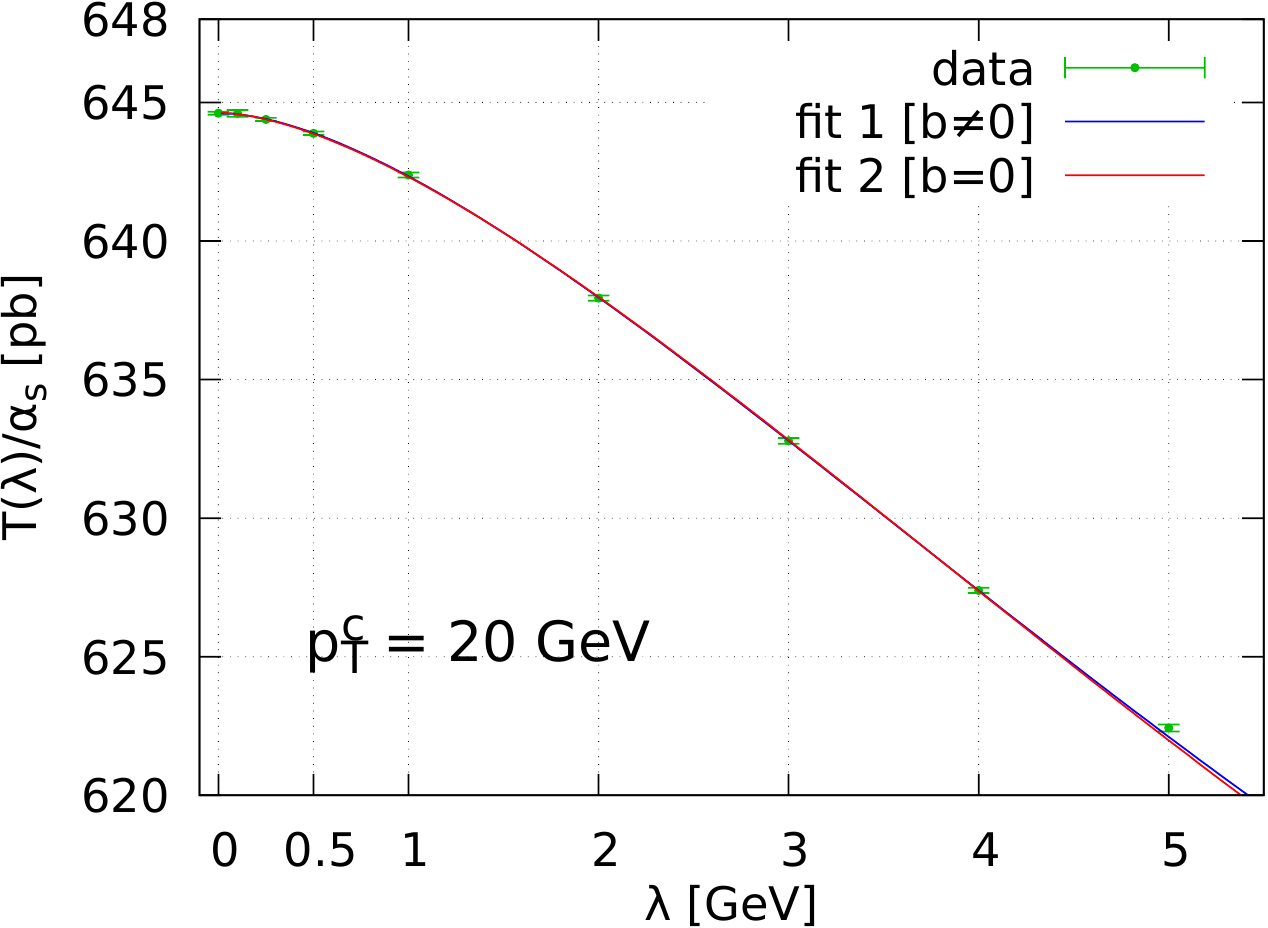}\hfill\includegraphics[width=0.48\textwidth]{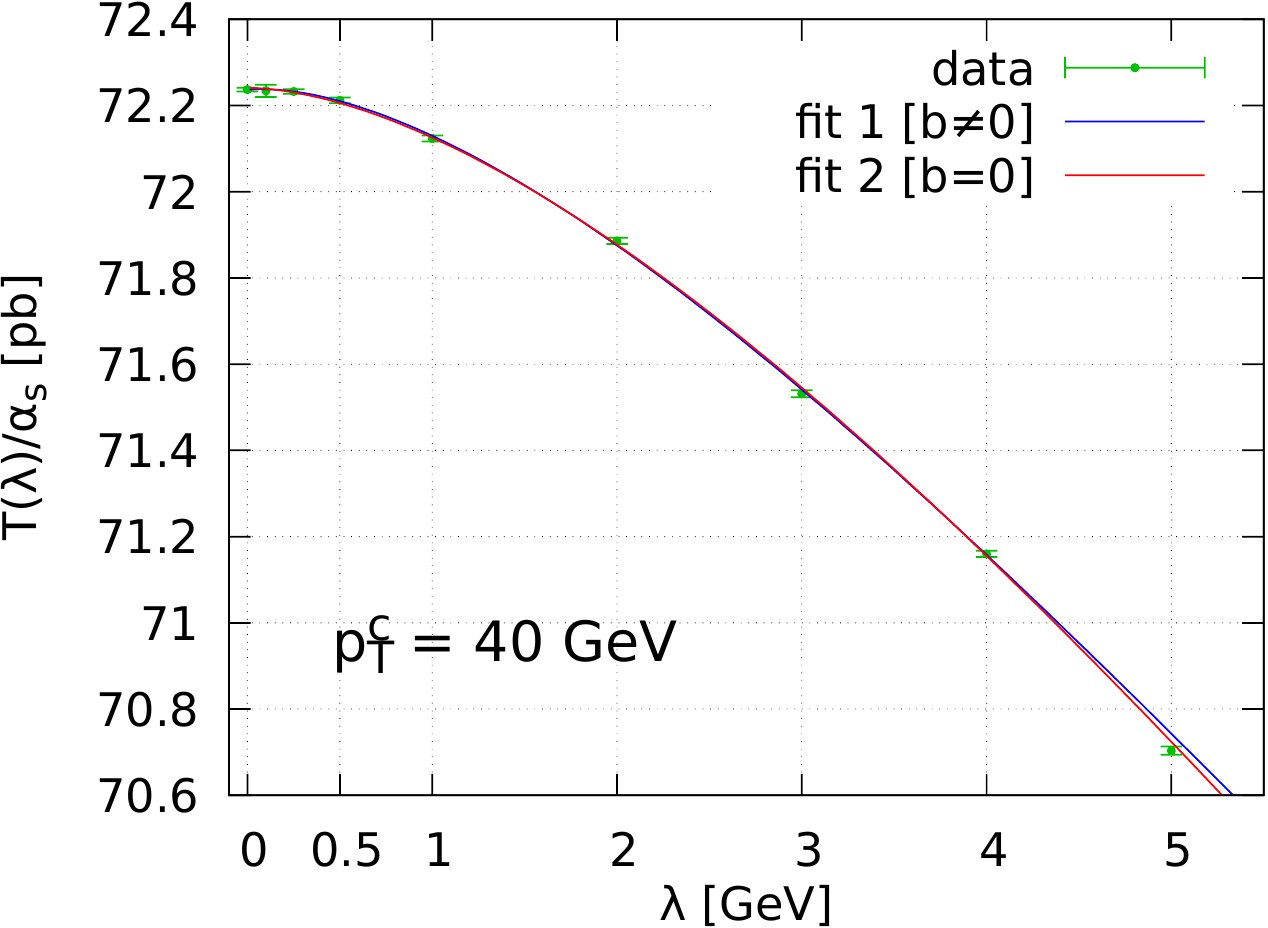}
   \caption{$T(\lambda)$ (defined in eq.~\eqref{eq:Tlambda}) as a function of the gluon mass $\lambda$
     for the total $Z$ production cross section with the cut $\ptz>\pt^c$, with $\pt^c=20\,$GeV (left)
     and $\pt^c=40\,$GeV (right). The green points are the results of our computations for several values
     of $\lambda$.
     The fit~1 and~2 lines are obtained with the fit function in eq.~\eqref{eq:T_fit_1}, where in fit~1
     all parameters are fitted, while in fit~2 the coefficient of the linear term $b$ is set to zero.  The point
     corresponding to $\lambda=5\,$GeV has not been included in the fits.}
   \label{fig:sigma_ptcut}
\end{figure}

To begin with, we show in fig.~\ref{fig:sigma_ptcut}
the result for the $T(\lambda)$ function defined in eq.~\eqref{eq:Tlambda}, associated to the cross section for the production of a $Z$ boson with
a transverse momentum larger than $20\,$GeV (fig.~\ref{fig:sigma_ptcut} on the left) and $40\,$GeV (fig.~\ref{fig:sigma_ptcut} on the right) as a function of the gluon mass $\lambda$.
In order to extract the slope around $\lambda=0$, which is responsible for linear renormalons (see eq.~\eqref{eq:Tslope}), we fit $T(\lambda)$
using the function
\begin{equation}
f(\lambda) = a \left[ 1+b\left(\frac{\lambda}{p_{\rm \scriptscriptstyle T}^c}\right)+c\left(\frac{\lambda}{p_{\rm \scriptscriptstyle T}^c}\right)^2 \log^2 \left(\frac{\lambda}{p_{\rm \scriptscriptstyle T}^c}\right) + d\left(\frac{\lambda}{p_{\rm \scriptscriptstyle T}^c}\right)^2\log\left(\frac{\lambda}{p_{\rm \scriptscriptstyle T}^c}\right)\right],
\label{eq:T_fit_1}
\end{equation}
where the inclusion of the single and double logarithmic terms are motivated by the findings in the Drell-Yan case~\cite{Beneke:1995pq,Dasgupta:1999zm}.
We neglected the point for $\lambda=5\,$~GeV in our fitting procedure, in order to increase the quality of the fit near $\lambda=0$.
We performed two fits, one including $b$ as fit parameter, and the other fixing it to 0,
in order to assess its impact on $T(\lambda)$.
In Tab.~\ref{tab:fit_pt} we report the results of the fits. We observe that the linear coefficient has a negligible impact on the fitting functions, its size is at least an order of magnitude smaller than the coefficient of the dominant quadratic term,
and its value is consistent with zero.
Thus, we find no evidence of the presence of a linear renormalon in the $Z$ boson transverse momentum distribution, and furthermore
we find that the value of the corresponding coefficient, if non-vanishing, is much smaller than the
coefficients of the quadratic terms.
\begin{table}[htb]
  \begin{center}
\resizebox{\textwidth}{!}
{   
\begin{tabular}{|c|c|c|c|}
\hline
\multicolumn{2}{|c|}{$p_{\rm \scriptscriptstyle T}^c=20\,$GeV}           & \multicolumn{2}{c|}{$p_{\rm \scriptscriptstyle T}^c=40\,$GeV}            \\ \hline
fit 1 & fit 2 & fit 1 & fit 2\\ \hline
$a=644.60\pm0.02$ & $a=644.63\pm0.02$ & $a=72.237\pm0.005$  & $a=72.241\pm0.004$  \\ \hline
$b=0.009\pm0.004$  & $b=0$                  & $b=0.024\pm0.017$ & $b=0$                  \\ \hline
$c=-0.063\pm0.008$ & $c=-0.047\pm0.004$ & $c=-0.11\pm0.06$ & $c=-0.028\pm0.021$ \\ \hline
$d=0.341\pm0.005$  & $d=0.341\pm0.007$  & $d=0.50\pm0.08$  & $d=0.59\pm0.05$  \\ \hline
$\chi^2/{\rm ndf}=0.12$  & $\chi^2/{\rm ndf}=0.23$  & $\chi^2/{\rm ndf}=1.13$   & $\chi^2/{\rm ndf}=1.36$  \\ \hline
\end{tabular}
}
\captionof{table}{Results of the fit of the  $T(\lambda)$ function, defined in eq.~\eqref{eq:Tlambda} and  illustrated in Fig.~\ref{fig:sigma_ptcut}.
The fit function is given in eq.~(\ref{eq:T_fit_1}).
In the first fit, corresponding to the blue lines in the figures, $b$ in unconstrained,
while in the second fit, corresponding to the red lines, $b$ has been set to 0. The last line corresponds to the associated reduced $\chi^2$.
}
\label{tab:fit_pt}
\end{center}
\end{table}

\begin{figure}[tb]
  \centering
    \includegraphics[width=0.48\textwidth]{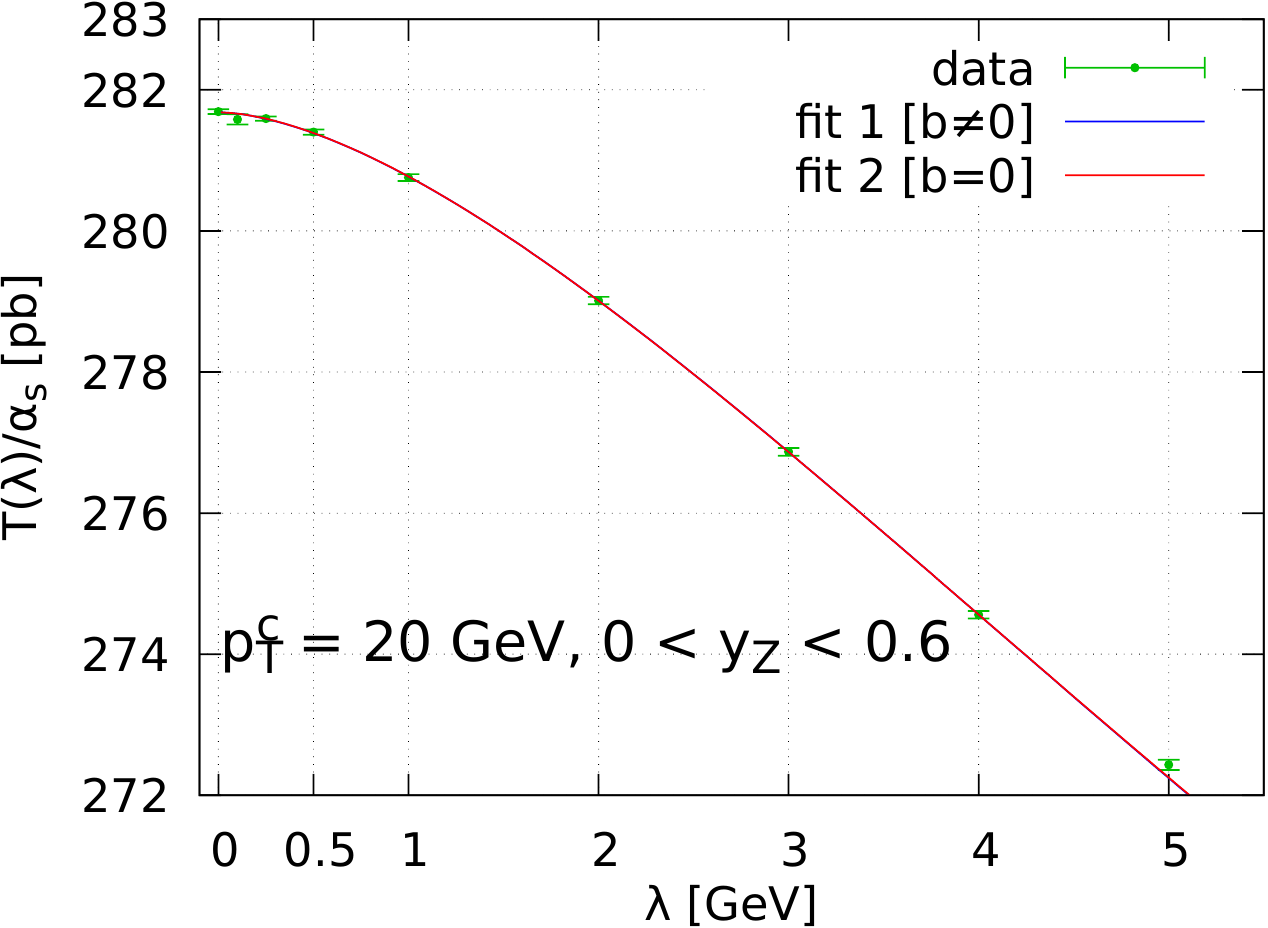}\hfill\includegraphics[width=0.48\textwidth]{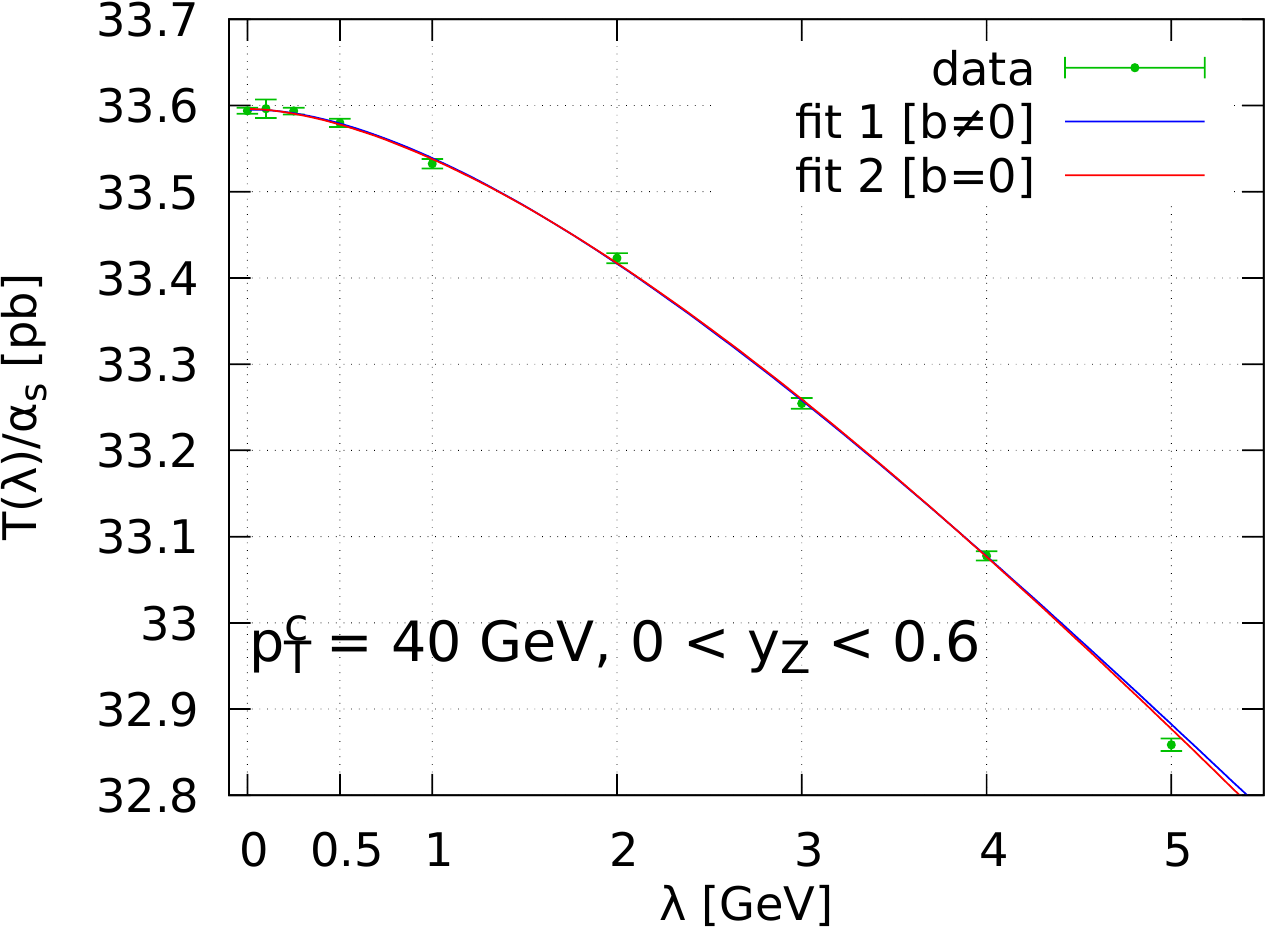}
   \caption{As in fig.~\ref{fig:sigma_ptcut}, supplemented with a cut on the $Z$ rapidity $0<y_{\rm \scriptscriptstyle Z}< y_c$,
     with $y_c=0.6$.}
   \label{fig:sigma_ptcut_y}
\end{figure}

We also performed a more exclusive analysis, imposing an additional cut over the rapidity of the $Z$ boson $\yz$,
besides the one over the transverse momentum. 
The results are shown in fig.~\ref{fig:sigma_ptcut_y}
and in Tab.~\ref{tab:fit_pt_y}.
\begin{table}[tb]
  \begin{center}
\resizebox{\textwidth}{!}
{   
\begin{tabular}{|c|c|c|c|}
\hline
\multicolumn{2}{|c|}{$p_{\rm \scriptscriptstyle T}^c=20\,$GeV}           & \multicolumn{2}{c|}{$p_{\rm \scriptscriptstyle T}^c=40\,$GeV}            \\ \hline
fit 1 & fit 2 & fit 1 & fit 2\\ \hline
$a=281.68\pm0.02$    & $a=281.68\pm 0.01$  & $a=33.595\pm 0.003$ & $a=33.596 \pm 0.002$ \\ \hline
$b=-0.001\pm0.009$     & $b=0$                & $b=0.015\pm 0.025$  & $b= 0$ \\ \hline
$c=-0.026\pm0.018$    & $c=-0.028 \pm 0.006$ & $c=-0.11\pm 0.09$   & $c=-0.06\pm 0.03$ \\ \hline
$d=0.35\pm0.01$     & $d=0.35 \pm 0.01$  & $d=0.49 \pm 0.11$    & $d=0.54\pm 0.06$  \\ \hline
$\chi^2/{\rm ndf}=0.39$  & $\chi^2/{\rm ndf}=0.32$  & $\chi^2/{\rm ndf}=0.89$   & $\chi^2/{\rm ndf}=0.77$  \\ \hline
\end{tabular}
}
\captionof{table}{Results of the fit of the  $T(\lambda)$ function defined in eq.~\eqref{eq:Tlambda}  illustrated in Fig.~\ref{fig:sigma_ptcut_y}. 
The fit function is given in eq.~\ref{eq:T_fit_1}.
The first fit corresponds to the blue lines, while in the second fit the linear coefficient has been set to 0 and corresponds to the red lines. The last line corresponds to the associated reduced $\chi^2$.
}
\label{tab:fit_pt_y}
\end{center}
\end{table}
Again we do not find numerical evidence of a linear sensitivity to $\lambda$, implying that the doubly differential distribution in rapidity and transverse momentum is free from linear renormalons.

By looking at the coefficients reported in Tabs.~\ref{tab:fit_pt} and~\ref{tab:fit_pt_y},
we notice that when we set $\pt^c=40$~GeV
instead of 20~GeV we encounter larger errors in the determination of
the coefficients $c$ and $d$, since the corresponding contributions
are suppressed by two powers of $\lambda/\pt^c$,
and thus their relative importance diminishes for higher cuts.

\section{Conclusions}\label{sec:Conclusions}
The current LHC physics demands high precision theoretical
predictions, and has promoted an unprecedented theoretical effort in
pushing perturbative calculations beyond next-to-leading order, and in
some cases even beyond the next-to-next-to-leading order. At the
current level of precision, possible non-perturbative effects that are
suppressed by a single power of the hard scale can sometimes be comparable or
larger in size than the current theoretical
uncertainties. Unfortunately, for collider physics observables we lack
a theory of even the most important non-perturbative corrections. This
is unlike others frameworks, including also heavy flavour physics,
where the existence of an operator product expansion allows us to
classify and parametrise non-perturbative effects.

Calculations of renormalon effects in Abelian models, using the so
called large-$b_0$ approximation, can provide a way to explore the
structure of non-perturbative effects in collider physics.  These
model calculations have helped in the past to investigate
the structure of renormalons in Drell-Yan
processes~\cite{Beneke:1995pq,Dasgupta:1999zm}, and have been used
recently in getting some insights on issues regarding the precision
measurements of the top mass~\cite{FerrarioRavasio:2018ubr}.  Related
methods have also contributed to a clarification of the structure of
linear renormalon effects in jet physics (see~\cite{Beneke:1998ui} and
references therein).

In this paper, we have used the large-$b_0$ approximation in order to
understand whether linear renormalon effects can yield linear power
corrections to the inclusive differential distribution for the
production of a vector boson at the LHC, in the regime where the
transverse momentum is safely in the perturbative region, and where
the resummation of transverse momentum logarithms is not needed. The
process we are considering is sufficiently complex, since it involves
gluon radiation both from the initial and final state, and entails
single and double logarithmic singularities that cancel when combining
the virtual and real corrections with the factorization of initial
state singularities.  We have chosen a process that mimics the $Z$
production in an Abelian model, but is such that, as in the full QCD
case, the soft radiation pattern is not azimuthally symmetric, and
thus, on intuitive ground, may be associated with non-perturbative
recoil effects that affect linearly the $Z$ transverse momentum.  We
find no evidence of linear power suppressed effects, and find instead
that the renormalon structure is well represented by quadratic terms
associated with some logarithmic enhancement.

Our numerical evidence gives us a useful indication, but, by its own nature,
can never be considered a solid proof of the absence of linear renormalons.  Nevertheless,
since we found that the coefficient of the linear renormalon is much smaller
than the coefficient of the quadratic one, and is consistent with zero,
we can conjecture that linear renormalons are absent for
the observables that we have considered.

Currently it seems that in all collider processes considered so far that involve massless
flavours, the large-$\bz$ approximation shows that
linear renormalons are absent in observables that are inclusive in
the production of coloured partons. This conclusion does not hold if massive quarks
are present, as reported in ref.~\cite{FerrarioRavasio:2018ubr}.
Further analytical work is needed in order to put these conjectures on more solid ground,
perhaps also shedding some light on the underlying mechanisms that lead to the
formation and cancellation of renormalon effects in hadron collider
physics.

\section*{Acknowledgments}
The authors want to thank Gavin P. Salam for interesting discussions
that have led us to consider the problem addressed in this work.  We
also thank Alexander Y. Huss, Simone Alioli, Martin Beneke, Mrinal Dasgupta, Pier Francesco Monni, Francesco
Tramontano and Giulia Zanderighi for useful comments on the
manuscript. P.N. acknowledges support from Fondazione Cariplo and
Regione Lombardia, grant 2017-2070, and from INFN.  S.F.R.'s work was
supported by the European Research Council (ERC) under the European
Union’s Horizon 2020 research and innovation programme (grant
agreement No. 788223, PanScales) and by the UK Science and Technology
Facilities Council (grant number ST/P001246/1).

\appendix
\section{Renormalon structure}\label{sec:AtanForm}
As far as the region of $\lambda<\muC$, that is the region relevant for IR renormalons,
a  linear term in $T(\lambda)$ will lead to the contribution
(see eq.~(\ref{eq:mainrenormform}))
\begin{equation}\label{eq:ren1}
  -\frac{1}{\bz\as}\left.\frac{\mathd T(\lambda)}{\mathd \lambda}\right|_{\lambda=0}
  \int_0^{\muC} \frac{\mathd \lambda}{\pi}
  \arctan \frac{\pi \bz \as}{1+\bz \as \log\frac{\lambda^2}{\muC^2}}
\end{equation}
Defining
\begin{equation}
  a\equiv \bz\as,
\end{equation}
the integral in eq.~(\ref{eq:ren1}) leads to the expression
\begin{eqnarray}
 \int_0^1 \frac{\mathd l}{\pi a}
\arctan \frac{\pi a}{1+a\log(l^2)}&=&
\frac{1}{\pi a} \arctan(\pi a) +\int_0^1 \mathd z \frac{\pi a z \cos(\pi z/2)-\sin(\pi z/2)}{1+(z\pi a)^2} \nonumber \\
&+&\frac{1}{\pi a} {\mathrm P}\int_0^\infty \mathd t \frac{\exp\left(-\frac{t}{2a}\right)}{1-t}-\frac{1}{a}\exp\left(-\frac{1}{2a}\right)\,, \label{eq:ren2}
\end{eqnarray}
where ${\mathrm P}$ indicates that the principal part of the integral should be taken.
The first two terms in eq.~(\ref{eq:ren2}) are obviously analytic in a neighbourhood of $a=0$.
The third term is a typical Borel representation of a power expansion with factorially growing coefficients,
resummed using a principal value prescription for handling the pole on the
real axis. If the pole is instead handled by moving the integration contour above or below the
real axis by a tiny amount, the imaginary part of the resulting integral is equal (in absolute value) to
the last term in the formula.
When replacing $a=\bz\as=1/\log{\muC^2/\Lambda^2}$, it leads to a linear power correction:
\begin{equation}
  \exp\left(-\frac{1}{2a}\right)=\frac{\Lambda}{\muC}\,.
\end{equation}
We thus see that the integral on the left-hand side of eq.~(\ref{eq:ren2}) cannot be cast in the form of a Borel sum
with principal value prescription, because of the
presence of the last term, whose origin can be traced back to the discontinuity of the arctangent
in the left-hand side integral when $\lambda=\lambda_{\rm\scriptscriptstyle  L}$, with
\begin{equation}
  \lambda_{\rm\scriptscriptstyle  L}=\muC\exp\left(-\frac{1}{2a}\right)
\end{equation}
In practical applications where one is interested in the value of the resummed expansion using the principal
value prescription, a suitable $\theta$ function is added to the arctangent integrand to make it continuous,
see ref.~\cite{Ball:1995ni}.
\bibliographystyle{JHEP}

\providecommand{\href}[2]{#2}\begingroup\raggedright\endgroup

\end{document}